\documentclass[pr12pt]{iopart}
\usepackage{bbold}
\usepackage{graphicx, caption}
\usepackage{iopams}
\expandafter\let\csname equation*\endcsname\relax
\expandafter\let\csname endequation*\endcsname\relax
\usepackage{amssymb}
\usepackage{amsmath}
\usepackage{empheq}
\usepackage{amsthm}
\usepackage{float}
\usepackage{siunitx}

\usepackage{color}
\usepackage{lineno}

\graphicspath{ {Figures/} }

\begin{document}


\title[Roton-like dispersion via polarisation change]{Roton-like dispersion via polarisation change for elastic wave energy control in graded delay-lines}

\author{L. Iorio \textsuperscript{1}, J. M. De Ponti \textsuperscript{1}, F. Maspero \textsuperscript{1} \& R. Ardito \textsuperscript{1}}

\ead{luca.iorio@polimi.it}
\address{$^1$Department of Civil and Environmental Engineering, Politecnico di Milano,
	Piazza Leonardo da Vinci, 32, 20133 Milano, Italy}

\begin{abstract} 
While roton dispersion relations had been restricted to correlated quantum systems at low temperature, recent works show the possibility of obtaining this unusual dispersion in acoustic and elastic metamaterials. Such phenomenon has been demonstrated in periodic structures by means of beyond-nearest-neighbor interactions, following the formulation firstly developed by Brillouin in the $'50s$. In this paper, we demonstrate both numerically and experimentally that beyond-nearest-neighbor connections are not a necessary condition to obtain this unusual dispersion relation in elasticity. Leveraging the intrinsic complexity of elastic systems supporting different types of waves, we demonstrate that mode locking can be applied to obtain roton dispersion, without the need of elastic or magnetic interactions between non nearest neighbors. Moreover, the combination of roton dispersion and rainbow physics enables spatial separation of the energy fluxes with positive and negative group velocity. 
\end{abstract}

\section{Introduction}
\label{sec:Intro}

Elastic and acoustic wave manipulation through the use of metamaterials and phononic crystals has been received increasing attention in the last two decades thanks to the exotic and non common dynamic properties of such structures.

While early designs were based on Bragg scattering or local resonance to create wide and low-frequency band-gaps \cite{Lemoult2011, Laude2015, Ardito2016, Achaoui2017, Krushynska2017, DAlessandro2020}, consideration is now transitioning to  more complete forms of wave control, including energy focusing and amplification \cite{Aguzzi2022,DePonti2019,DePonti2020}, wave filtering \cite{de2021elastic}, beam-splitting \cite{Makwana2018}, robust signal communication \cite{Miniaci2018, Cha2018} and localization \cite{Xia2020} using concepts from topological insulators \cite{Chen2018, Nassar2020,ChaplainTopological2020}. The possibility to control wave energy is always related to the design of properly engineered dispersion relations, with the concurrent creation of band-gaps and peculiar eigenstates. In topologically based metamaterials,  protected edge (or interfacial) surface states are achieved by leveraging broken symmetries in time \cite{Wang2015, Souslov2017, Shankar2022} or space \cite{Miniaci2018,Mousavi2015, Mousavi2015}, emulating the Quantum Hall (QHE) or Quantum Spin Hall (QSHE) effects. In graded-based systems, band-gaps are spatially controlled to emulate lensing or rainbow effects for signal amplification \cite{DePonti2020, DePonti2021b} or energy conversion \cite{colombi16a, Chaplain2020Umklapp}. Broadly speaking, a large number of studies have been devoted to shaping the wave dispersion curves, creating phononic, locally resonant or topological band-gaps, often combined together to obtain unusual behaviors. 

Very recently, theoretical \cite{WegenerTeo2021} and experimental studies \cite{WegenerExp2021} in acoustics and elasticity show the possibility of tailoring the lowest dispersion branch to achieve a so-called \textit{roton-like} dispersion. 
Generally the term, most certainly for elastic waves, has been used to describe dispersion relations of a material that shows two points of zero-group velocity (one is a relative maximum and the other a relative minimum). The term originated in quantum physics from the peculiar dispersion of the quasi-particle Helium-4, that shows a decrease in momentum when the energy of the particle is increased \cite{Landau1941, Feynman1956}. This is due to the generation of a relative rotation of the particles that effectively reduces the momentum of each particle in a specific energy region \cite{Glyde1990,Godfrin2021}. The term has been then expanded to the realm of designed crystals or periodic metamaterials, even though the involved phenomena are physically very different. Such unusual dispersion has been obtained using chiral crystals with  hybridization of the microrotational and translational modes \cite{Kishine2020} or by means of non nearest-neighbor interactions. The idea of using $L^{th}$ neighbour interactions to get multiple wavelengths at a given frequency, i.e. dispersion curves with multiple maxima and minima, was firstly developed by Brillouin \cite{Brillouin1946} in the $'50s$, with a complete discussion about the distance of interaction in lumped systems. Such ideas have been recently revitalized, designing engineered structures with  non-nearest-neighbor interactions by means of elastic connections \cite{WegenerTeo2021, WegenerExp2021} or thanks to electrostatic or magnetic forces \cite{Carcaterra2019, Rezaei2021, Lacarbonara2020}. While both approaches enable the effective design of a so-called roton-like dispersion curve, they imply complex manufacturing techniques. Specifically, non-nearest-neighbor interactions based on elastic connections require three dimensional setting to avoid overlap in two dimensions \cite{WegenerTeo2021, WegenerExp2021}. This can be a serious limitation for many applications, e.g. in Micro-electromechanical systems (MEMS) \cite{Corigliano2017} or large structures which cannot rely on 3D-printing technologies. On the other hand, the mathematical model of the second approach is more difficult, given the intrinsic non-linear nature of the  magnetic or electrostatic interaction, and manufacturing issues can arise in creating complex elasto-magnetic structures.

On this paper, we propose alternative and simpler structures, which can be easily manufactured in two-dimensional settings without the need of complex arrangements of elastic connections or non-linear magnetic forces. 

The method we propose is general for elastic system, and allows to obtain the same dispersion relation characteristics as the works described above, but not relying on coupling between cells far apart. The underlying physics is based on the peculiar properties of elastic systems, i.e. their capability to naturally support different types of waves, each one with a specific wavelength, polarisation and velocity. Since an elastic medium simultaneously support different propagating modes with different wave speeds, a coupling between them automatically activates non-nearest-neighbors without the necessity of extra elastic links or forces. 

This approach relies on a physical phenomenon known in physics as mode locking \cite{Mace2012}, and recently proposed to achieve rainbow trapping \cite{ChaplainDelineating2020} and selective mode conversion \cite{DePonti2021, DePonti2022} in elastic waveguides. By breaking symmetries along the wave propagation direction, a strong coupling between different modes is achieved granting, at a given frequency, waves with equal and opposite group velocity. Specifically, we design an arrangement of resonators able to couple three different types of wave: torsional ($T$), flexural in-plane ($F_{ip}$) and flexural out-of-plane ($F_{op}$), each one with a specific wavelength $\lambda$, polarisation $p$, phase $v_{ph}$ and group $v_g$ velocity. While torsional waves are non-dispersive, flexural waves are dispersive and, for the problem considered here, the torsional wavespeed $v_T$ is lower than the flexural one $v_T$ ($v_T<v_F$).

\begin{figure}[t!]

\includegraphics[width = 1\textwidth]{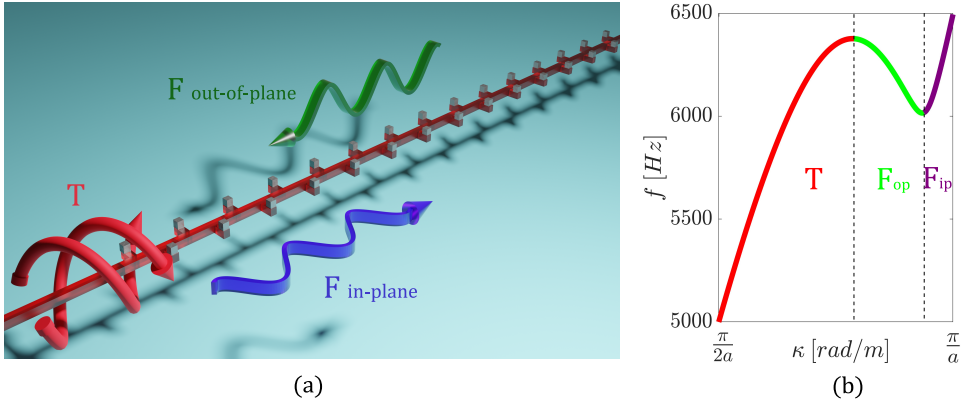}
\caption{(a) Rendering of the roton-like structure and the waves it supports and couples. The torsional wave, denoted with $T$ (red spiral arrows), enters the metamaterial and it is initially mode converted into a counterpropagating out-of-plane flexural wave ($F_{op}$). In a second step, the out-of-plane flexural wave, with negative group velocity, couples while traversing the array with an in-plane flexural wave ($F_{ip}$). The wave coupling is obtained thanks to the brake in symmetry of the resonators along the wave propagation direction. Leveraging on this phenomenon, a dispersion relation with a local maximum and minimum within the First Brillouin Zone (FBZ) can be obtained. Such dispersion, named as roton-like, is depicted in (b), were each type of wave is colour coated.  By inspecting the dispersion curve we notice the opposite group velocity of the out-of-plane flexural wave with respect to the torsional wave and the in-plane flexural wave, suggesting, at given frequency, competing energy fluxes. 
}
    \label{fig:intro}
\end{figure}

The concept of coupling and the consequent roton-like dispersion is schematically reported in Fig. \ref{fig:intro}, where the periodic structure that couples the three modes is shown. The modes are also schematically depicted in Fig. \ref{fig:intro} (a) (red: torsional, violet: in-plane flexural, green: out-of-plane flexural), while Fig. \ref{fig:intro} (b) shows the dispersion relation of the elementary cell colour coated with respect to the mode described. The coupling between $T$, $F_{op}$ and $F_{ip}$ is achieved by placing asymmetric resonators along the waveguide. By doing so, at a given frequency, three different wavevectors are excited, two of which with positive group velocity ($T$ and $F_{ip}$), and one with negative group velocity ($F_{op}$). The physics involved gives rise to an energy flux competition between the different wave modes, redirecting energy with different polarisation and wave speed. However, as usually noticed  in locking-based wave systems, the sudden change of the wavenumber induced by the periodic structure provides strong reflection. This mechanism, which is a physical consequence of the conservation of the crystal momentum, can be overcome by using graded structures \cite{DePonti2021, DePonti2022}, i.e by adiabatically modifying the wavenumber transformation along space. For this reason, in the last section we combine roton-based and rainbow physics \cite{Tsakmakidis2007}. To do so, we gradually modify the geometry of the cell, and consequentially the dispersion curve, so that the relative maximum of the dispersion becomes the minimum in a subsequent cell. This type of dispersion engineering shows that the wave is heavily localised in the metamaterial, while also being efficiently converted into all the possible available modes that make up the roton-like dispersion. The result is a graded roton-like structure that is able to slow down a torsional incoming wave, convert it multiple times and finally give back a propagating in-plane flexural mode. 
The numerical and experimental results presented herein demonstrate how a simple structure can be used to confine and mode convert elastic waves, expanding the range of possibilities in the context of wave manipulation and control, with implications of technological relevance for applications involving mechanical vibrations, such as nondestructive evaluation, ultrasonic imaging, and energy harvesting. Moreover, the simultaneous coupling between waves with opposite energy fluxes paves the way for the development of effective delay-lines which can be used for sensing or signal amplification.

\begin{figure}[t!]
    \centering
    \includegraphics[width = 1\textwidth]{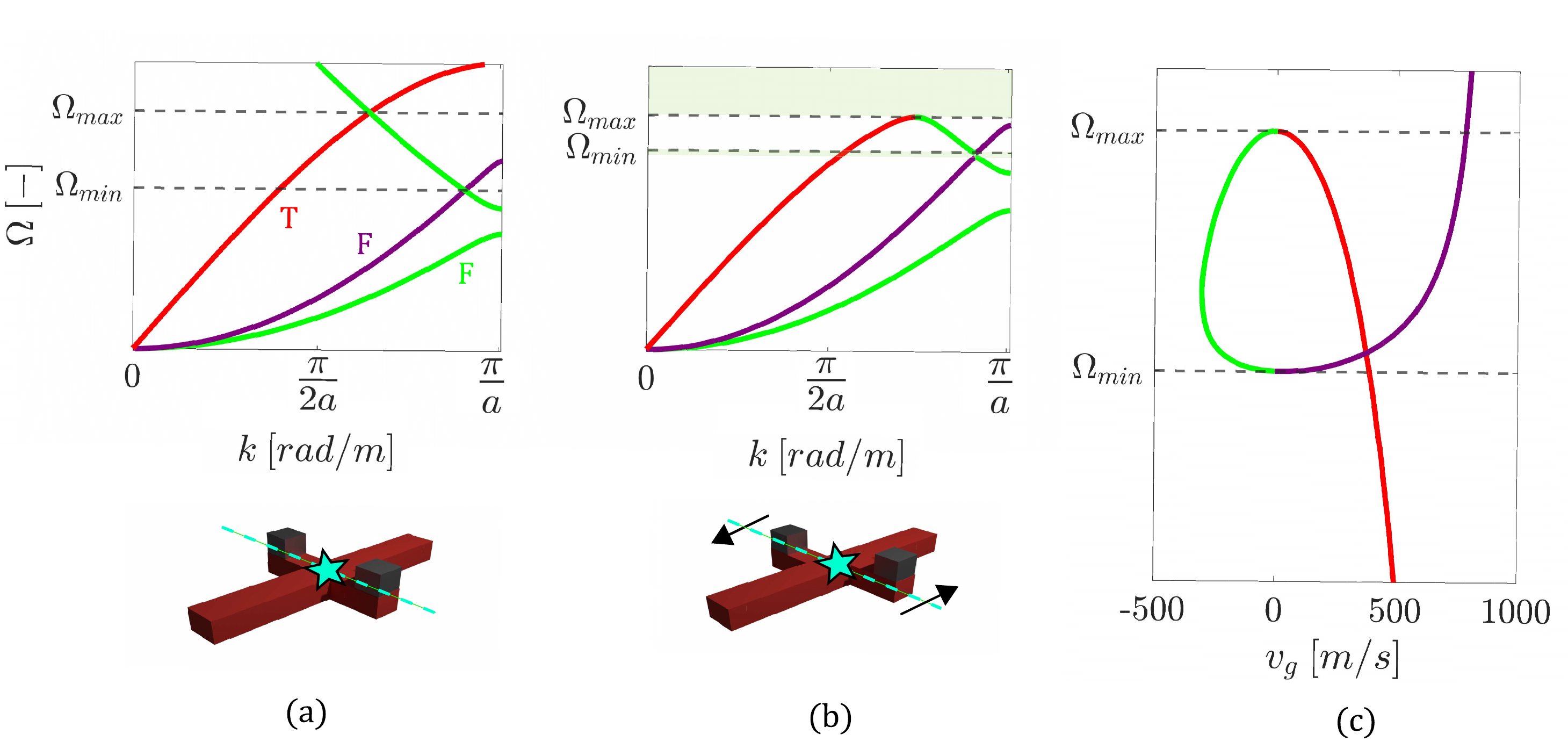}
    \caption{Dispersion relation for the symmetric (degenerate) (a) and the shifted (lifted degeneracy) (b) cell. When the cell is mirror symmetric with respect to the wave propagation direction (a), band crossings occurs between $T$, $F_{op}$ and $F_{ip}$ (marked with red, green and violet respectively). In such case, the eigenmodes are orthogonal, i.e. completely decoupled. By breaking the mirror symmetry the degeneracy is lifted, resulting in the creation of band-gaps and mode coupling. Leveraging on this phenomenon, a roton-like dispersion relation can be obtained, showing positive, negative and zero group velocities (c). This is the case depicted in the third branch of the dispersion in (b). 
    }
    \label{fig:02}
\end{figure}

\section{Designing the dispersion through mode locking}
\label{sec:LineArrays}
As stated above, the design implemented that grants the roton-like dispersion is obtained through the coupling between counter-propagating modes that results in the required maximum and minimum of the dispersion curve.
This is possible thanks to the locking coupling mechanism that enables the interaction between two counter propagating modes that initially suffer accidental degeneracy \cite{Mace2012, DePonti2021}. By modifying the waveguide structure, it is possible to lift the degeneracy, create a small band-gap, and enable the interaction and the transfer of energy between the two modes in the vicinity of the band gap. If this mechanism is engineered twice on the same dispersion curve and by correctly stacking the modes in the right order and at the right frequency, the ending result is the roton-like dispersion relation. To obtain this result, the local group velocity of the three modes has to be engineered so that one of the three modes has a negative group velocity in the degeneracy region. After the coupling, the dispersion will have two points of zero-group velocity (one that is a local maximum and one that is a local minimum) as it is shown in Fig. \ref{fig:intro} (b). To engineer the waveguide properties, it is first necessary to find and analyse a unit cell that is then periodically repeated to create an infinite waveguide; this is amenable to Floquet-Bloch theory and one extracts dispersion relations that relate the frequency to the Bloch wavenumber; the geometry we start with, Fig. \ref{fig:02} (a), has points of accidental degeneracy between three distinct branches of the dispersion relation. These points are necessary for the creation of band-gaps generated by lifting the degeneracy through the coupling of the wave modes. Fig. \ref{fig:02} (b) reports the unit cell and the dispersion curves when the degeneracy is lifted. The twofold and simultaneous lift in degeneracy is given by a relative shift in position of the lateral resonators that, from a symmetric positioning in the cell in the degenerate case, get staggered. The final cell, the one used from this moment onward, is obtained by maximising the shift in relative position of the resonators that end up giving the cell a glide symmetry, that grants an elimination of the Bragg band gap. Moreover, Fig. \ref{fig:02} (c) also reports the change of group velocity ($v_g = \frac{\partial \omega}{\partial k}$) of the roton-like dispersion branch with respect to frequency, colour coated depending on the relative mode considered. The phase velocity ($v_{ph}= \frac{\omega}{k}$) is always greater then 0, while the group velocity can be positive, zero and even negative in certain frequency ranges. 
The dispersion curves obtained up to now are all computed numerically by means of a Finite Element Model (FEM)  using COMSOL Multiphysics\textsuperscript{\textregistered} and imposing Bloch-Floquet boundary conditions on the elementary cell that makes up the metamaterial beam. 

\begin{figure}[t!]
    \centering
    \includegraphics[width = 0.85\textwidth]{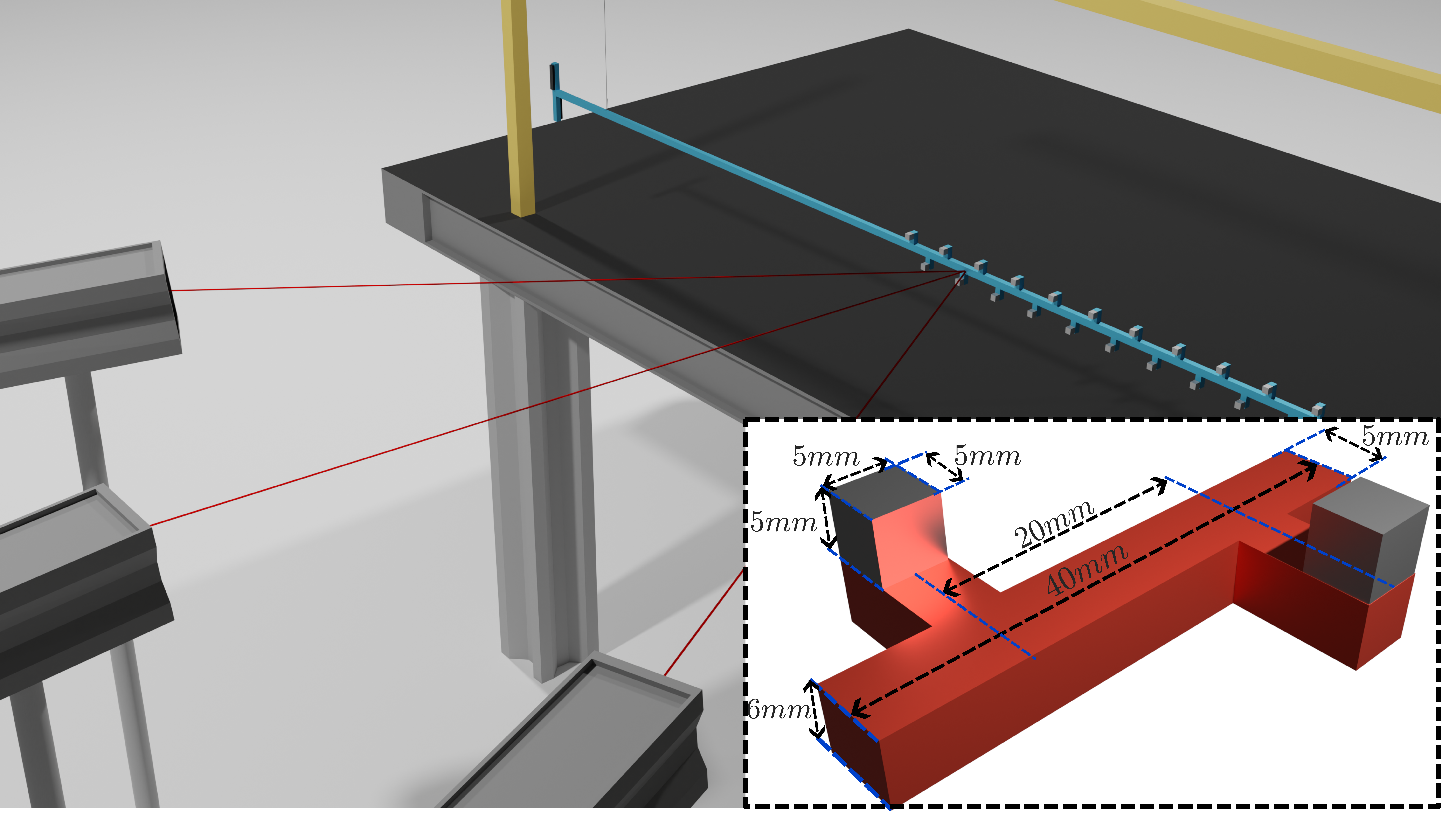}
    \caption{Rendering of the experimental setup: the roton structure is hanged with wires, the input is given by a couple of resonators at the ledt end of the beam which are actuated out-of-phase using piezoelectric patches. Three heads of the Scanning Laser Doppler Vibrometer evaluate the velocity field inside the structure. On the bottom right the enlargement of the elementary cell's geometry that composes the periodic roton-like structure is reported.}
    \label{fig:03}
\end{figure}

\begin{figure}[t!]
    \centering
    \includegraphics[width = 1\textwidth]{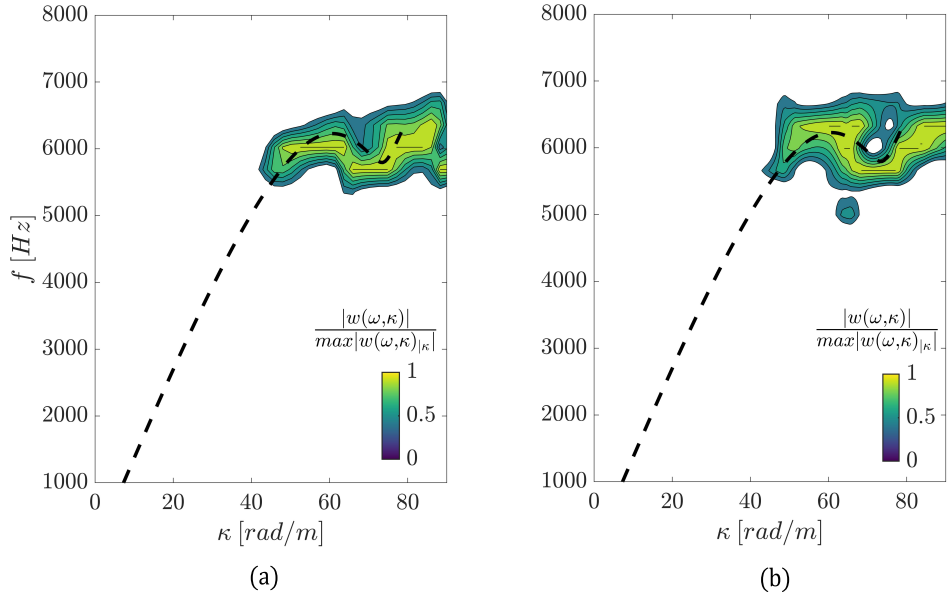}
    \caption{Numerical and experimental dispersion relation of the roton-like structure. (a) Numerical result obtained from time domain analysis and (b) experimental results using the same input over the constructed geometry. The dashed black line is the numerical dispersion relation obtained through Bloch-Floquet boundary condition applied to the cell.}
    \label{fig:04}
\end{figure}

We validate numerically and experimentally the roton-like dispersion by inspecting, in time domain, the wave propagation along the periodic waveguide. The system is composed of twelve periodic cells so that a clear indication over the values of the wavevector in the periodic array can be obtained. A detailed description of the periodic structure and the experimental setup is shown in Fig. \ref{fig:03}. The structure is made primarily of a steel beam, obtained from a plate using water jet cutting technology. The thickness is 5 mm while its width is 6 mm. The resonators, each 10 mm long and 5 mm wide, are positioned at a distance of 20 mm. As for the tip mass implemented, for simplicity we opt for neodymium-iron-boron magnets of size 5x5x5 mm. The magnetic field is used to easily attach the tip masses, while its force doesn't interfere with the dynamic response of the system. The magnets are too far apart to interact between themselves and the magnetic field lines are deviated into the underlying steel structure. For this reason, we numerically neglect magnetic forces and consider the magnets perfectly attached to the resonators.  The advantage of this structure is that it is very easy to manufacture with respect to current literature solutions \cite{WegenerTeo2021, WegenerExp2021}, because it doesn't involve the use of complex three dimensional structures to couple far away cells, or magnetic systems.  

For the experimental setup, we suspend the beam on a frame structure by means of elastic cables that do not affect the dynamics of the system. The excitation signal is provided through a $KEYSIGHT$ $33500B$ waveform generator which synchronously starts with the measurement system. We use a narrowband source $V(t) = V_0w(n)\sin(2\pi f_c)$ with amplitude  $V_0= 80$ ${\rm V}$, Hann window $w(n)$, central frequency $f_c= 6000$ Hz and time duration $0.5$ ms. The input is applied in the form of a torsional wave excitation; this is obtained experimentally by placing on one side of the waveguide a couple of resonators which are actuated out-of-phase using piezoelectric patches. The velocity field of the structure is measured using a Polytec 3D Scanning Laser Doppler Vibrometer (SLDV), that is able to measure and separate the 3D velocity field on the waveguide inside the array of resonators.
To corroborate the roton-like dispersion of the structure, we inspect the time-domain wavefield when a narrowband input torsional wave is applied. We then evaluate the dispersion by applying a two-dimensional (2D) Fast Fourier Transform (FFT) both in time and in space.
It can be noticed that the numerical (Fig. \ref{fig:04} (a))  and experimental (Fig. \ref{fig:04}) (b) two-dimensional FFT of the wavefiled are in good agreement with the expected roton-like dispersion. Specifically, we notice the peculiar maximum and minimum within the FBZ, together with positive and negative group velocity regions in the wavenumber space.

\section{Adiabatic roton-like delay lines}
\begin{figure}[t!]
    \centering
    \includegraphics[width = 1\textwidth]{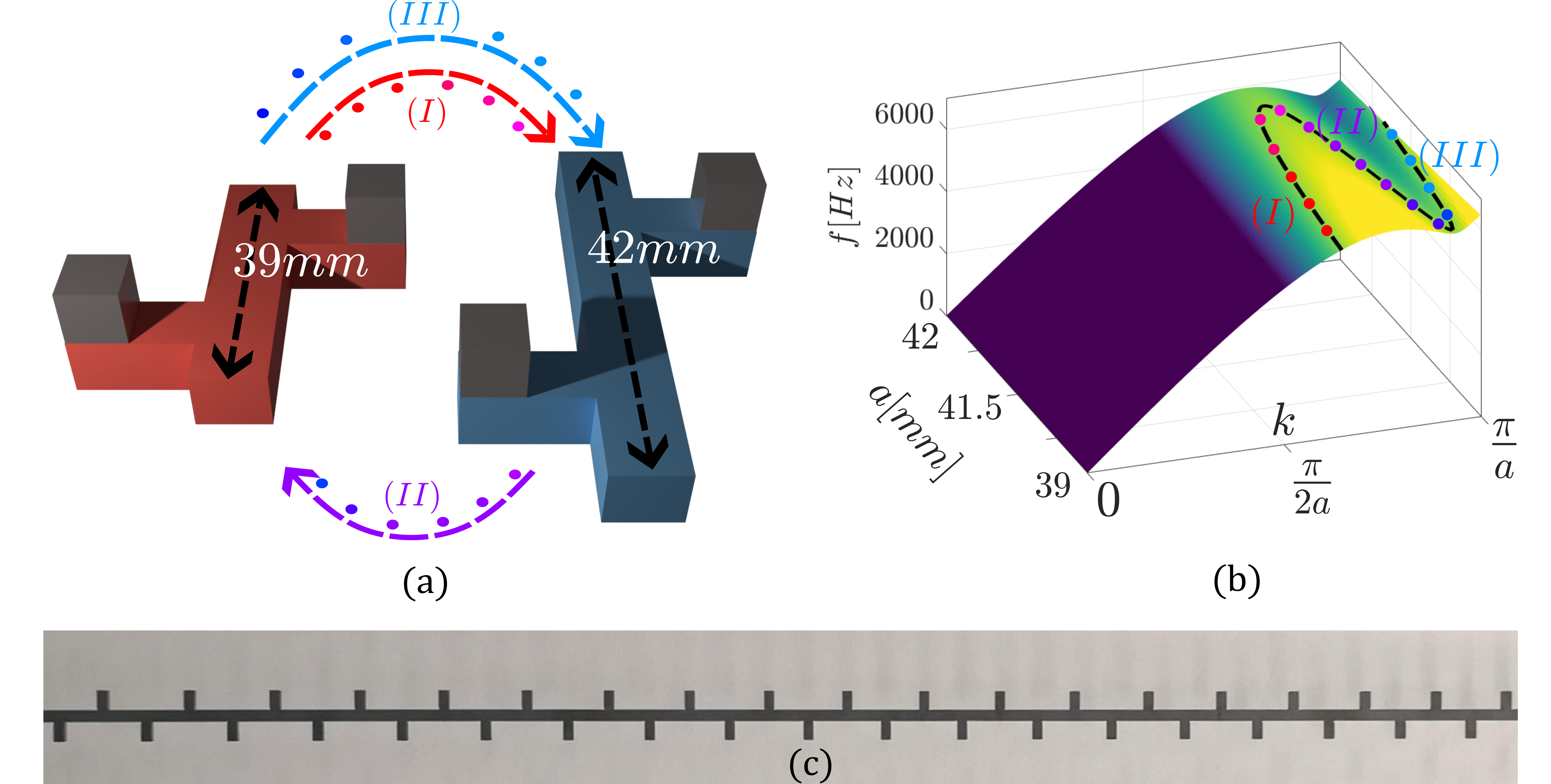}
    \caption{(a) Geometry of the last and first cell of the graded roton-like array. The grading is obtained by varying the cell length, from 39 mm to 42 mm, keeping the resonators distanced half the cell length. (b) 3D representation of the dispersion relation in exam with respect to the normalised wavevector, the cell dimension and the frequency. The black dashed line is the path the wave follows during its propagation in the graded array. (I), (II), (III) Denote the the paths that the different waves, with different group velocities, take while traversing the system and being slowed down and mode converted in the array.
    In (a) the paths are represented by the coloured dashed lines and the arrows show from which cell the waves start and arrive at. (c) Image of one of the specimens used for the experiments.}
    \label{fig:05}
\end{figure}

Having obtained the roton-like dispersion is not effective for wave control given that the actual coupling between the modes doesn't allow one mode (the input wave) to efficiently and effectively follow all the positive and negative parts of the dispersion branch. In other words, a torsional mode sent to the structure would be mostly backscattered when near the local maximum. This is a consequence of the conservation of crystal momentum, which induces reflections for fast and abrupt wavenumber transformations.
For this reason, we opt for designing a graded roton-like structure. The aim is to implement an almost adiabatic grading of the system in the wavenumber space \cite{DePonti2021, DePonti2022} so that the wave can be more efficiently converted into all the three modes. This mechanism of smooth change in the wavevector and so in homogenised properties, eases the wave into the array, given that the mismatch in wavenumbers between the plain beam waveguide and the array is greatly reduced.
To engineer the graded structure, the smooth change in properties is obtained by changing the distance between the resonators, and so the unit cells dimensions. Specifically, the geometry and the material properties of the cells remain the same (resonators lengths, width, tip mass), but the actual length of the cell is modified by changing the distance between the resonators. In the array, that is made of 25 cells, the cells are progressively enlarged starting from a length of 39 mm and ending the system with a length of 42 mm.
This modifies the resulting roton-like dispersion as shown in Fig. \ref{fig:05}. Here, the geometrical dimensions of the first and last unit cells of the graded structure are reported (Fig. \ref{fig:05} (a)), together with a three-dimensional dispersion  (Fig. \ref{fig:05} (b)). It shows how the dispersion changes respectively with the enlargement of the cell. A level line is also present at the central frequency at which the analyses is performed. This is done to highlight the path that the wave is going to follow both in the real space and in the reciprocal space. The input is a torsionally polarised propagating wave, that enters the array. Then, inside the array (path (I)), the wave is slowed down and mode converted into a back propagating out-of-plane flexural mode. This happens progressively and it is over before it reaches the last cells of the adiabatic array. When converted, the back propagating out-of-plane flexural mode retraces the entire array (path (II)) and it is progressively converted into a propagating in-plane flexural mode near the beginning of the array. This new mode then propagates (path (III)) in the same direction as the first torsional wave from which it was generated. This final mode traverses again the entire array, finally exiting on the other side of the graded roton-like structure.

\begin{figure}[p]
    \centering
    \includegraphics[width = 1\textwidth]{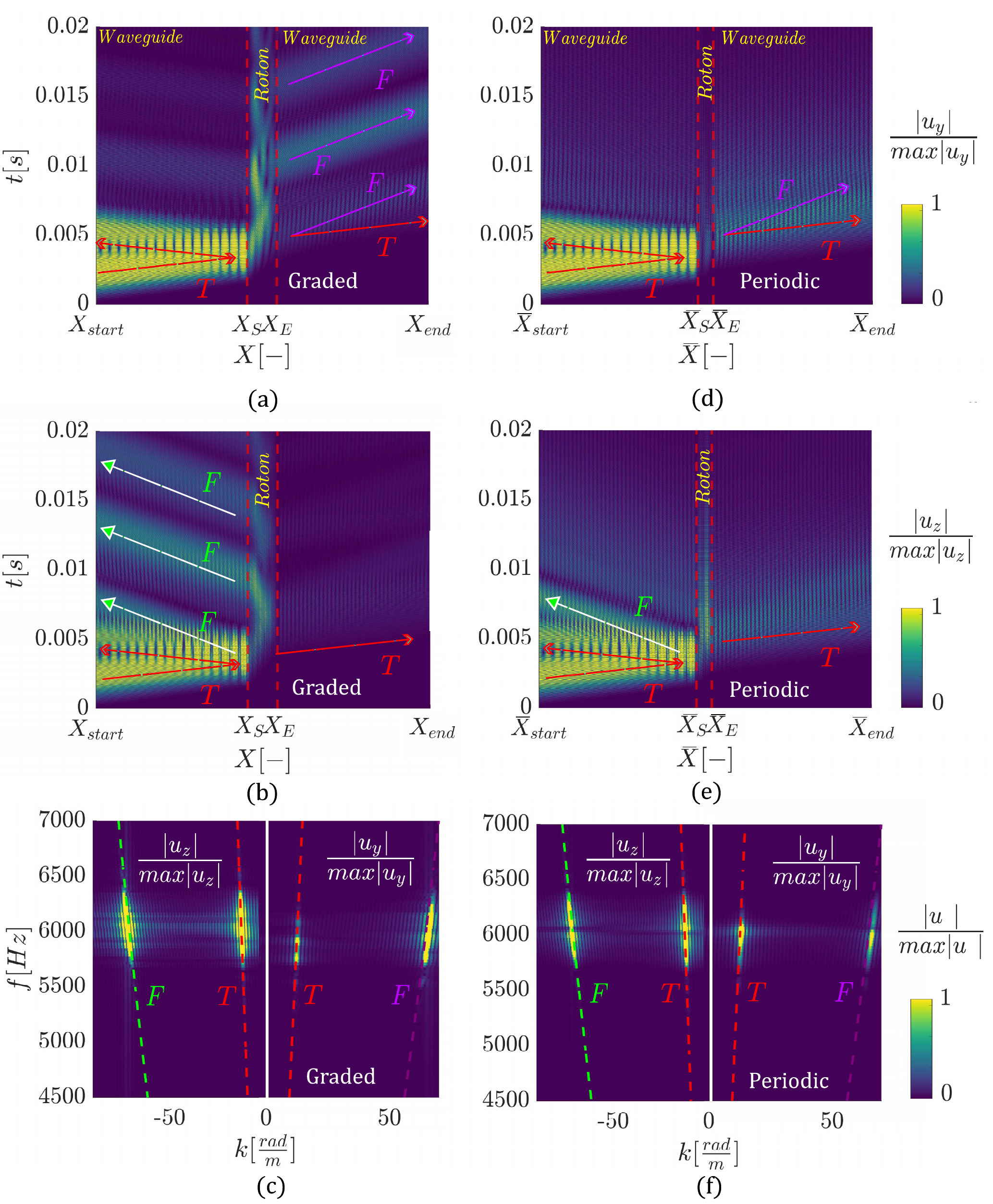}
    \caption{Waterfall plots and spatiotemporal FFTs from time domain analyses over both the graded (a, b, c) and the periodic (d, e, f) structures. The structures are connected to an elastic waveguide on both sides, with absorbing boundary conditions. Results are obtained by taking a line of points on one edge of the waveguide to show both the input and output waves. The displacement along y-direction is inspected to catch the in-plane behaviour ($F_{ip}$), while the displacement along z-direction is measured to describe the out-of plane response ($F_{op}$). By comparing the waterfall plots (a,d) - (b,e), a stronger conversion of flexural waves is observed in the graded structure. Similarly, the FFT from the graded structure (c) shows a stronger wavenumber contribution of both $F_{op}$ and $F_{ip}$ with respect to the periodic case (f).
    }
    \label{fig:06}
\end{figure}

This hypothesis of enhancement of multiple wave conversion, coupled with the idea of creating an efficient delay line that converts a propagating torsion into a propagating flexural wave is studied both numerically and experimentally. From the numerical side, we perform an implicit time-domain analysis over the graded and periodic roton-like structures. Fig. \ref{fig:06} shows the results in the form of waterfall plots and spatiotemporal FFTs both in time and space. The arrays are both excited with the same input wave: a torsional wave centered at 6050 Hz. Both arrays are connected before and after to plain beams with the same dimension and properties of the main beam that composes the backbone of the roton-like arrays. At both ends of these beams, Absorbing Layers using Increasing Damping (ALID) \cite{RAJAGOPAL201230} are applied to avoid reflections that are generated either by the input wave or by the output modes of the array.

The numerical waterfall plots are reported in Fig. \ref{fig:06} for the graded (a,b) and periodic (d,e) structures. We show the wavefield along y and z direction, to be able to catch both the in-plane and out-of-plane flexural modes. Moreover, to identify both torsional and flexural contributions, we opt for a line of points located on the edge of the waveguide; this choice is motivated by the negligible contribution of torsional waves on the displacement field along the central beam axis. Starting with the analysis of the graded roton array, we see that, for the in-plane motion (Fig. \ref{fig:06} (a)), the input torsion is heavily scattered by the array. The wave that enters the array is deeply confined inside of it in both time and space. The outputs given by the array are then twofold: there is a small component of torsional wave that exits the array (it is not converted by the adiabatic roton, but its intensity is very small), while a second component (and the most relevant one) is an in-plane flexural wave that, after a significant delay, exits the array. This component is the one we are most interested in, because it is the one hypothesised and described also in the previous paragraph about what waves we expect to see, where, when and in which directions they propagate. The distinction between the different modes is possible thanks to the very different wave speeds. Moving to the analysis of the out-of-plane motion (Fig. \ref{fig:06} (b)) we again see that the torsion enters the array, the wave is confined in the array and then it is backscattered after some time as an out-of plane flexural mode into the first waveguide. This continuous radiation of the wave signal is a typical behaviour of rainbow structures, denoted in acoustics as time spreading of the reflected pulse \cite{Cebrecos2014}. The structure is able to confine and radiate energy at specific time intervals, providing an effective wave conversion. The existence of propagating and reflected torsional waves is attributed to the grading, which is not gentle enough \cite{DePonti2021}. For this reason, the second conversion (out-of plane to in-plane) is not 100\% efficient. We expect better results (higher conversion efficiency) if the graded structure includes more cells. Fig. \ref{fig:06} (c) shows the same results of \ref{fig:06} (a,b) but in the reciprocal space instead of the real space. To correctly detect the propagating $F_{ip}$ and reflected $F_{op}$, we decide to focus on the positive wavenumber from the y displacement and the negative wavenumber from the z displacement.
A strong amount of torsional wave is converted into flexural waves both in-plane and out-of-plane. 
To conclude the discussion over the results of the graded array, we can say that the system acts according to the prediction of a double rainbow trapping effect \cite{DePonti2021}, given the two points of zero group velocity in the reciprocal space and the graded effect. The wave conversion is efficient and both flexural waves that are generated are first confined in the system and then exit the array after a significant time delay.

We now move to the results for the periodic roton-like array. For what concerns the in-plane solution (Fig. \ref{fig:06} (d)) the input torsion is mostly backscattered or it traverses the system somewhat delayed. Moreover, a small flexural component can be seen after the array but it is very limited in intensity. A similar behaviour can be noticed for the out-of-plane solution (Fig. \ref{fig:06} (e)), where most of the torsion is reflected and a small amount of the wave is converted. Finally, we can see that the wave is not localised in the array contrary to the graded case. Moving to the reciprocal space (Fig. \ref{fig:06} (f)), we notice that the periodic structure is able to transfer a lower amount of energy on the flexural modes with respect to the graded system. In conclusion, we notice that, without the graded array, the propagating in-plane flexural mode can not be efficiently generated, and that the wave is not strongly delayed.

\begin{figure}[p]
    \centering
    \includegraphics[width = 1\textwidth]{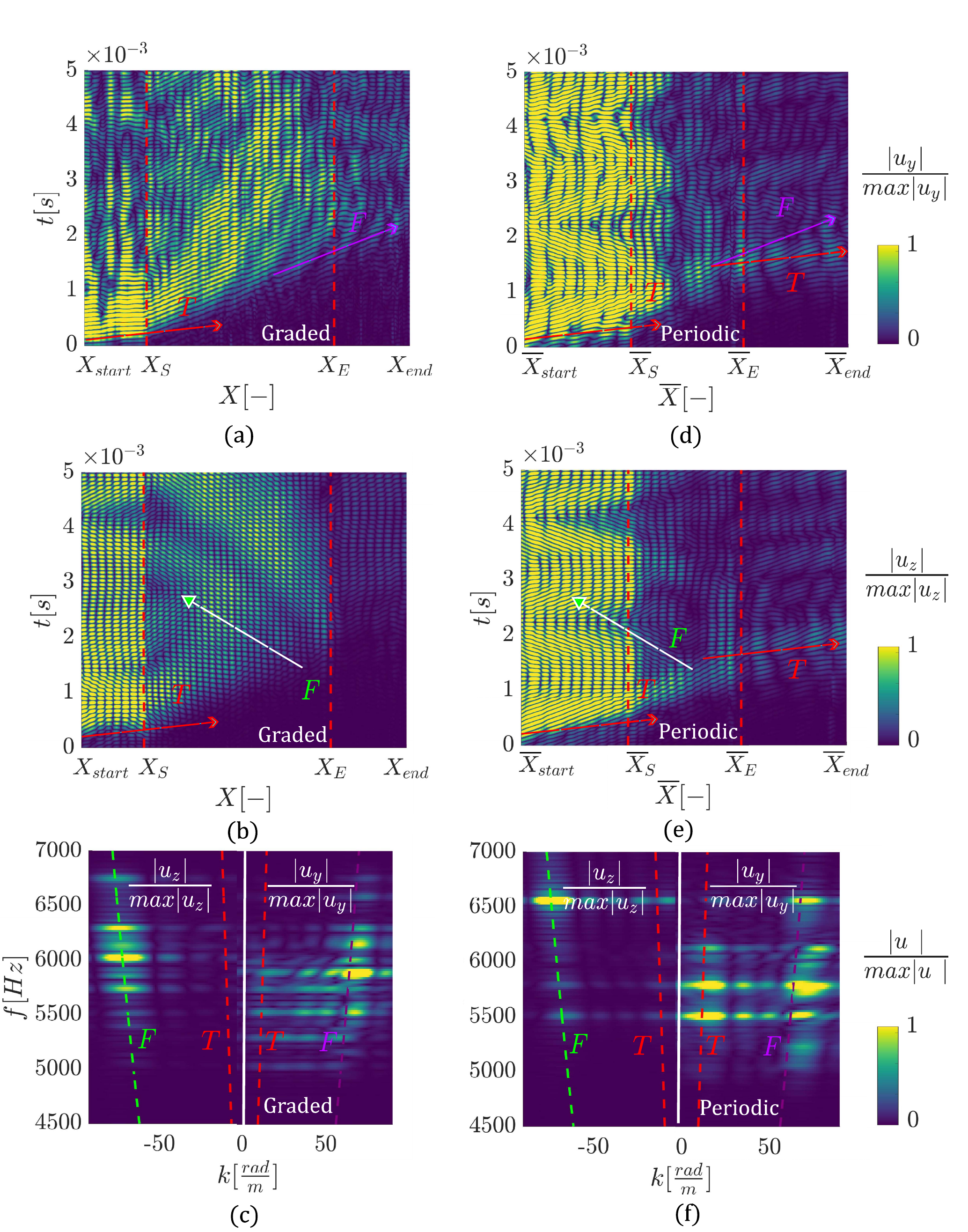}
    \caption{Experimental waterfall plots and spatiotemporal FFTs over both the graded (a, b, c) and the periodic (d, e, f) structures. Differently with respect to the numerical results, both structures are of finite lengths, thus allowing for spurious reflections. Results are obtained by taking a line of points on the waveguide axis; for this reason, torsional waves cannot be fully detected looking at the z direction. As for the numerical results, a stronger conversion of flexural waves is observed for the adiabatic structure. This is particularly visible from the FFTs, where we notice a higher contribution of $F_{ip}$ and $F_{op}$ in the graded structure (c) with respect to the periodic one (f). We also outline the presence of stronger torsional contributions in the periodic structure (f) which confirms its low performance in terms of wave conversion.}
    \label{fig:07}
\end{figure}

The same structures are tested experimentally using the setup reported in Fig. \ref{fig:03}. The only difference with respect to the numerical simulations is the length of the waveguide before and after the array, which is not infinite (for obvious reasons). This limitation has two implications: i) the input signal, generated by the piezoelectric patches, has to be shorter to properly see a propagating wave and so the excited frequencies are more spread out;  ii) since we want to scan the entire length of the structure, a lower resolution along the beam width is expected. For this reason, the points are taken on the central line axis of the beam, resulting in a less clear indication of the input torsional wave. The input is provided by two out-of-phase piezoelectric patches positioned over two lateral resonators at the beginning of the waveguide (same configuration of Fig. \ref{fig:03}). The results are reported in Fig. \ref{fig:07} in the form of waterfall plots and FFTs. We can clearly distinguish the difference between the graded roton-like array and the periodic one. Even though the results are not as clear as the numerical ones, mainly because of the absence of absorbing boundary conditions and the presence of imperfections in the specimen, the overall behaviour expected is maintained. The graded structure better converts the input torsion into the two flexural modes, while the periodic array is less able to do so and we see a major contribution given by the torsional mode, both in the FFT space and the waterfall plots.

\section{Conclusions and Perspectives}
A roton-like dispersion relation for an elastic waveguide was engineered and studied both numerically and experimentally. The change in wave polarisation is the key ingredient to obtain such unusual dispersion without  implementing  connections between non-nearest neighbors. The creation of a local maximum and minimum is given by the coupling through wave locking of different wave modes that, thanks to the staggered resonators, interact and exchange energy. The resulting effect is that waves with different wavelengths and group velocity propagate simultaneously in opposite directions inside the array. The roton-like nature of the dispersion relation has been validated both numerically and experimentally. Furthermore, we also demonstrated how to enhance the energy exchange between the modes in the roton-like array by spatially modifying the parameters of the elementary cell, i.e. leveraging the rainbow effect. By introducing a grading, a double rainbow trapping effect was created, allowing for a wave confinement in the array for extended periods of time. The focus though was mainly on studying the resulting waves that are radiated from the array after extended time intervals. The most interesting one is the in-plane flexural wave that is generated after a strong delay and is then mostly the only wave that is outputted on the other side of the rainbow array. The experiments also showed the importance of the graded array, and reproduced, with some differences given by the complications of having finite structures, the results shown numerically. In the end, the graded roton-like waveguide we propose has the features of not having complicated fabrication methods to obtain the dispersion requested and it can be easily tuned to the needed frequencies. The results show promising avenues both for structures that implement mode coupling as a tool for dispersion engineering and the synergy that these systems have with graded arrays. Signal control can be achieved to create delay lines and elastic energy localisation zones, useful tools for energy harvesting applications.

\section{Acknowledgements}
The support of the H2020 FET-proactive Metamaterial Enabled Vibration Energy Harvesting (MetaVEH) project under Grant Agreement No. 952039 is acknowledged. The authors wish to thank D. Cavazzi for the prototype realization and Prof. F. Braghin for the experimental tests.

\end{document}